\documentclass[prl,amssymb,twocolumn,tightenlines,showpacs,superscriptaddress]{revtex4-1}

\usepackage{amsmath}
\usepackage{physics}
\usepackage{graphicx}
\usepackage{dcolumn}
\usepackage{bm}

\usepackage{hyperref}
\hypersetup{colorlinks, citecolor=blue, linkcolor=blue, urlcolor=blue}

\usepackage{editing}
\usepackage{units}
\usepackage{textcomp}
\usepackage{xspace}
\usepackage{amsmath}
\usepackage{amssymb}
\usepackage{gensymb}
\usepackage{booktabs} 
\usepackage{colortbl} 
\usepackage{xcolor} 
\usepackage{xfrac}

\newcommand{\Ca}{\ensuremath{^{40}{\rm Ca}^+}\xspace}

\newchange{}{DarkGreen}{OrangeRed}%

\changeOnOff

\begin{document}

\preprint{APS/123-QED}

\title{Coherent Control of the Rotational Degree of Freedom of a Two-Ion Coulomb Crystal}

\author{Erik Urban}
 \affiliation{Physics Department, University of California, Berkeley}
\author{Neil Glikin}%
 \affiliation{Physics Department, University of California, Berkeley}
\author{Sara Mouradian}%
 \affiliation{Physics Department, University of California, Berkeley}
\author{Kai Krimmel}
 \affiliation{Helmholtz-Institut Mainz, Mainz, Germany}%
 \affiliation{QUANTUM, Institut f¨ur Physik, Johannes Gutenberg-Universit¨at Mainz, Mainz, Germany}
\author{Boerge Hemmerling}
 \affiliation{Department of Physics and Astronomy, University of California, Riverside} 
\author{Hartmut Haeffner}
 \affiliation{Physics Department, University of California, Berkeley}%

\date{\today}

\begin{abstract}
We demonstrate the preparation and coherent control of the angular momentum state of a two-ion crystal. The ions are prepared with an average angular momentum of $7780\hbar$ freely rotating at 100~kHz in a circularly symmetric potential, allowing us to address rotational sidebands. By coherently exciting these motional sidebands, we create superpositions of states separated by up to four angular momentum quanta. Ramsey experiments show the expected dephasing of the superposition which is dependent on the number of quanta separating the states. These results demonstrate coherent control of a collective motional state described as a quantum rotor in trapped ions. Moreover, our work offers an expansion of the utility of trapped ions for quantum simulation, interferometry, and sensing.

\end{abstract}

\maketitle

Coherent control of the collective motion of trapped-ion Coulomb crystals is fundamental to their versatility as a platform for quantum control. Historically, ions have been trapped in a linear chain with motional modes well-modeled as harmonic oscillators~\cite{Wineland1998,Leibfried2003review,Haeffner2008review}. However, the utility of trapped ions for quantum simulation, fundamental physics, and sensing can be further expanded by access to and control over collective motional modes with more complex dynamics.

For example, consider the motion of a quantum rotor. The energy spectrum of the angular momentum eigenstates is quadratic in quantum number, increasing the complexity possible in Hamiltonian engineering and enabling the simulation of rotational dynamics of diatomic molecules~\cite{Judson1990,Shen1991,Yuan2011}. Moreover, the periodic boundary conditions of angular momentum states enable fundamentally new operations such as the deterministic coherent exchange of the ions' wave functions~\cite{Roos2017} and the ability to create interferometry geometries new to ions~\cite{Campbell2017, Barrett2014}. Rotational states also have fundamental physics applications in Aharanov-Bohm style experiments because of their enclosed area~\cite{Weigert1995,Noguchi2014} and in observing Hawking radiation in acoustic analogs of black holes~\cite{Horstmann2010a,Horstmann2011}. Finally, rotor states interact with noise in interesting ways due to their extended size and spatial symmetry which could have applications in metrology and sensing~\cite{Sindelka2006,Lenef1997,Yang2017, Cryan2011, Fogedby2018,Pelegri2018}.

In this paper, we describe control over the angular momentum eigenstates of a pair of \Ca ions in a cylindrically symmetric surface-electrode Paul trap. Through classical preparation of a high angular momentum state, we are able to spectrally separate densely spaced angular momentum transitions into groups according to the number of angular momentum quanta involved in the transition. We optically drive a group of sidebands and observe Rabi oscillations which demonstrate that the transitions are coherent and match our presented theory. Additionally, Ramsey experiments show the expected dephasing due to the non-linearity of the energy of rotor states, a phenomenon not present in harmonic oscillator superpositions. The capability to produce coherent superpositions of angular momentum states not only expands the toolbox of trapped ion experiments but also brings the capabilities of trapped ions to the study of molecular dynamics, periodic systems, and rotational sensing.

\begin{figure*}
    \centering
    \includegraphics[scale=.68]{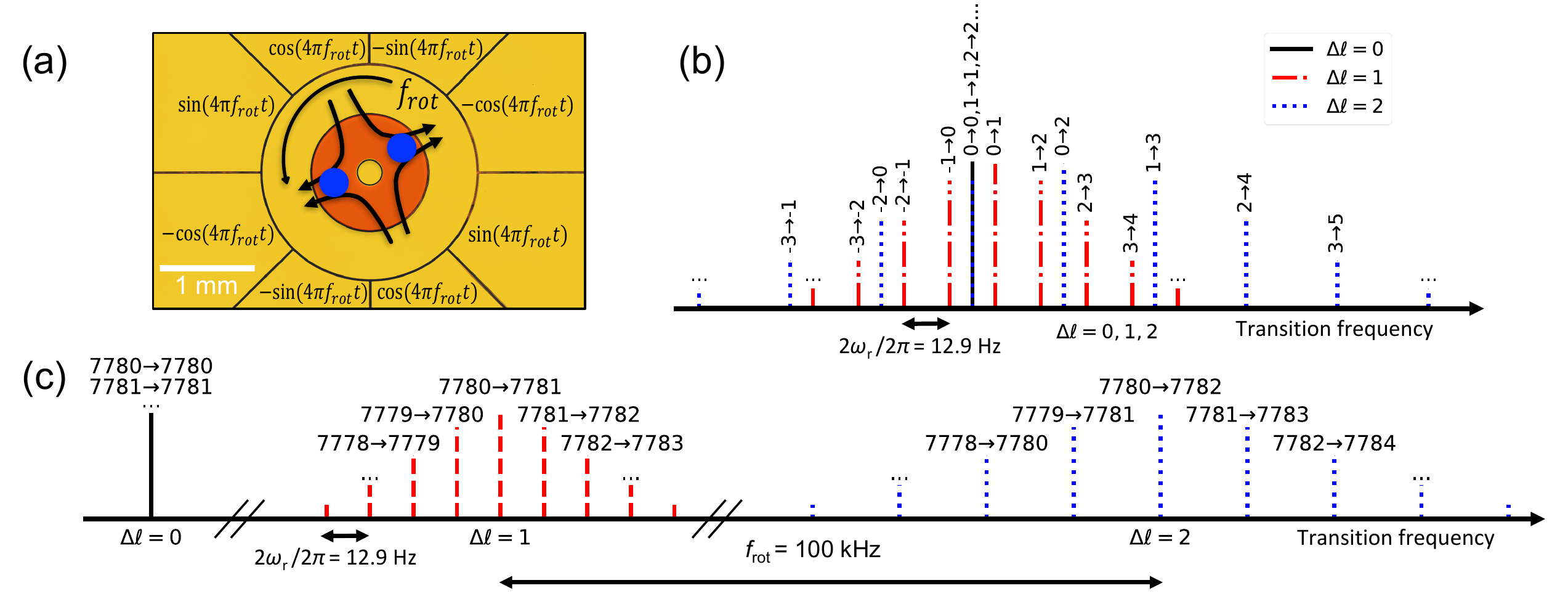}
    \caption{(a) False color, optical image of the trap center. The center electrode and the outer ring are grounded while RF is applied to the second circular electrode (red). Voltages used to create the rotating quadrupole, the fields they produce, and the relative orientation of the ions (not to scale) are shown on the figure. (b) Frequency of angular momentum transitions relative to the carrier for a stationary ion ring, each separated by only a few Hz. Transitions are labeled by their initial and final state quantum numbers. Heights are proportional to occupation of the initial state quantum number for $\sigma_\ell=2$. (c) Frequency of angular momentum transitions relative to the carrier for an ion ring rotating at 100~kHz. Groups of lines are separated by $f_\textrm{rot}$ with width in frequency space of $4\omega_{\rm r}\sigma_\ell\Delta\ell$. }
    \label{fig:spectrum}
\end{figure*}

Our rotor is produced by loading two ions into a trap with a single radio-frequency (RF) null. In this potential, the mutual Coulomb repulsion of the ions repels them from the center and they form a small ring. The point potential is created 184~$\mu$m above the surface of the trap shown in Fig.~\ref{fig:spectrum}a by applying RF to the second circular electrode and grounding the other two circular electrodes~\cite{Wang2015}. Eight compensation electrodes surrounding the RF and ground electrodes are used to compensate both dipole and quadrupole stray electric fields at the trapping location, thereby creating a cylindrically symmetric, 3D harmonic potential~\cite{Li2017}. If all in-plane DC quadrupole potentials are compensated, the two horizontal trap frequencies are degenerate, $\omega_x$, $\omega_y = 2\pi\times845$~kHz, while the vertical trap frequency $\omega_z \approx 2\omega_x$. 

In this potential, two ions form a crystal with an equilibrium radius of $r_e=\left[\frac{e^2}{16\pi\epsilon_0}\frac{1}{m\omega_x^2}\right]^{1/3} = 3.13~\mu$m where $m$ is the mass of \Ca. Such a crystal has no preferred angular orientation within the $xy$-plane. Ignoring micromotion and separating out the common mode harmonic motion, the in-plane Hamiltonian of two ions in cylindrical coordinates is 
\begin{equation}
    H = -\frac{\hbar^2}{m}\left(\pdv[2]{\rho} + \frac{1}{\rho}\pdv{\rho} + \frac{1}{\rho^2}\pdv[2]{\phi}\right) + \frac34 m\omega_x^2(\rho-r_e)^2
\end{equation}

\noindent
where $\rho$ and $\phi$ are the radial and angular polar coordinates, respectively, in the $xy$-plane. Within the approximation of harmonic confinement in the radial direction, this is equivalent to the Hamiltonian of a trapped diatomic molecule with its motion confined to two dimensions~\cite{McQuarrie1997}, where the interatomic potential is formed by the harmonic confinement of the trap field and the Coulomb repulsion. This Hamiltonian is approximately a rigid rotor with radius $r_e$ in $\phi$, and a harmonic oscillator with frequency $\sqrt{3}\omega_x$ in $\rho-r_e$. These modes of motion are coupled to each other by a Coriolis term which is small if the rotational frequency $2\pi f_\textrm{rot} \ll \omega_x$ where $f_\textrm{rot}$ is the angular frequency of the rotor. In this approximation, the angular eigenfunctions are $Y_\ell(\phi) = \frac{1}{\sqrt{2\pi}}e^{i\ell\phi}$ with energies $E_{\ell} = \frac{\hbar^2\ell^2}{4mr_e^2}= \hbar\omega_{\rm r}\ell^2$, where $\ell$ is the quantum number labeling the angular momentum states and $\omega_{\rm r}=\frac{\hbar}{4mr_e^2}=2\pi\times 6.43$~Hz is the fundamental frequency scale of the rotor. 

The motional modes of trapped \Ca ions are coherently controlled by optically addressing the transition from the electronic ground state, $\ket{S} \equiv~^{2}$S$_{1/2}$, to a long-lived, metastable state, $\ket{D} \equiv~^2$D$_{5/2}$, on the spectral sidebands created at the motional eigenfrequencies of the ion crystal~\cite{Cirac1994b,Monroe1995a,Roos1999}. To allow coherent control, these sidebands must be spectrally narrow and isolated. While a harmonic oscillator mode creates sidebands at integer multiples of its characteristic frequency, a rigid rotor mode has a nonlinear energy spectrum and thus spectral sideband addressing is more complicated. To coherently manipulate the ion crystal's angular momentum states, the motional sideband corresponding to the transition from $|\ell\rangle$ to $|\ell+\Delta\ell\rangle$ must be resolved in frequency space from all other $\Delta\ell$ transitions for all initial $\ket{\ell}$ which have appreciable population. In a thermal state, angular momentum states are populated as a Gaussian centered about $\ell_0 = 0$ parameterized by a standard deviation $\sigma_\ell$ related to the temperature. At the Doppler limit (0.52~mK for \Ca), $\sigma_\ell = 920$. Fig.~\ref{fig:spectrum}b shows the expected transition frequencies of only  $\Delta\ell = \{0,1,2\}$ rotational transitions weighted by the initial state's occupation, under the condition $\ell_0=0$ and $\sigma_\ell=2$ for illustration. Individual lines of various $\Delta\ell$ transitions are interspersed amongst each other. To address an individual line, the stability of the energy difference between states and the laser linewidth would both need to be well under 10~Hz, which is prohibitively narrow. 

If instead the population of the angular momentum states is centered at a sufficiently large angular momentum  $\hbar\ell_0$, corresponding to a rotation frequency of $2\pi f_\text{rot}=2\omega_{\rm r}\ell_0$, transitions spectrally separate from each other grouped by their order $\Delta\ell$, as shown schematically in Fig.~\ref{fig:spectrum}c. While the individual lines remain separated by only $2\Delta\ell\omega_{\rm r} \approx \Delta\ell\times2\pi\times12.9$~Hz, each group as a whole becomes individually addressable as long as the separation between these groups $2\pi f_\text{rot}$ is significantly greater than the group's width in frequency space $4\omega_{\rm r}\sigma_\ell\Delta\ell$. For this reason, the parameter $\sigma_\ell$ determines both how narrow each transition is and which transitions can be resolved. We use $\sigma_\ell$ to characterize the width of the state in angular momentum space throughout this paper. 

To achieve control over individual $\Delta\ell$ transitions we prepare a rotating ($f_\textrm{rot} = 100$~kHz, $\ell_0 \approx 7780$) and cold (low $\sigma_\ell = 46$) two-ion crystal in three stages. First, we break the angular symmetry of the trapping potential with a static, in-plane quadrupole generated by voltages $V\cos{(\alpha_0+\alpha_i)}$ applied to each DC electrode (shown in Fig.~\ref{fig:spectrum}a) where $\alpha_i$ is shifted by $\pi/4$ relative to its counterclockwise neighboring electrode. This pins the orientation of the ion crystal and creates an in-plane tilt mode of up to $\omega_\text{tilt} = 2\pi\times280$~kHz, the orientation of which is controllable by $\alpha_0$. Next, with the ions pinned, the phase $\alpha_0$ of the in-plane quadrupole is ramped to accelerate the angular orientation of the quadrupole to a final angular velocity of 100~kHz in a time of 50~$\mu$s. The voltages are sourced by a single arbitrary waveform generator applying an accelerating sine wave whose phase is shifted appropriately for each electrode with a custom-built circuit. Finally, after reaching the target angular velocity, the quadrupole continues to rotate at the final speed while the amplitude is reduced linearly to zero over 1~ms. When the rotating quadrupole is completely turned off, the ions continue rotating due to conservation of angular momentum, and now do so in the desired symmetric potential. Experimentally, the quoted spin-up and release times give us the narrowest $\sigma_\ell$ within our technical limitations. It is critical to cool the in-plane tilt mode to the ground state prior to spin-up, since the thermal occupation of the tilt mode directly maps onto a Gaussian distribution of angular momentum states during the release process.
We find that $\sigma_\ell \approx 400$ when only Doppler cooling is performed on the pinned crystal which only separates transitions up to $\Delta\ell = 4$ by two standard deviations. With the addition of resolved sideband cooling of the in-plane tilt mode, $\sigma_\ell$ is reduced to 46, allowing us to potentially resolve transitions up to $\Delta\ell = 42$.

To control the quantum angular momentum state of the ion crystal, we address a group of rotational sidebands of the $\ket{S}\rightarrow\ket{D}$ transition (729~nm), all corresponding to the same change in angular momentum state, $\Delta\ell$. In the rotating wave approximation, the relative coupling strength between the states $|\ell\rangle$ and $|\ell+\Delta\ell\rangle$ is given by

\begin{equation} \label{eq:mat_el}
	\mel{\ell+\Delta\ell}{e^{ik_xx}}{\ell} = J_{\Delta\ell}(k_xr_e)
\end{equation}

\noindent
where $k_x=k\cos\theta$ is the projection of the laser's wavevector in the plane of the rotor for an angle $\theta$ between the wavevector and the rotor plane and $J_{\Delta\ell}$ is a Bessel function of the first kind of order $\Delta\ell$. The maximum sideband order $\Delta\ell_\textrm{max}$ at which there is still significant sideband coupling strength is proportional to the Doppler shift observed by the laser from the rotating ions, and roughly given by $\Delta \ell_\textrm{max}\approx k_x r_e \approx 27$ if $\theta=0$. In order to consolidate the oscillator strength into a few transitions, the excitation laser is aligned nearly perpendicular to the rotation plane to reduce $\Delta\ell_\textrm{max}$. 

\begin{figure}
    \centering
    \includegraphics[scale=.4]{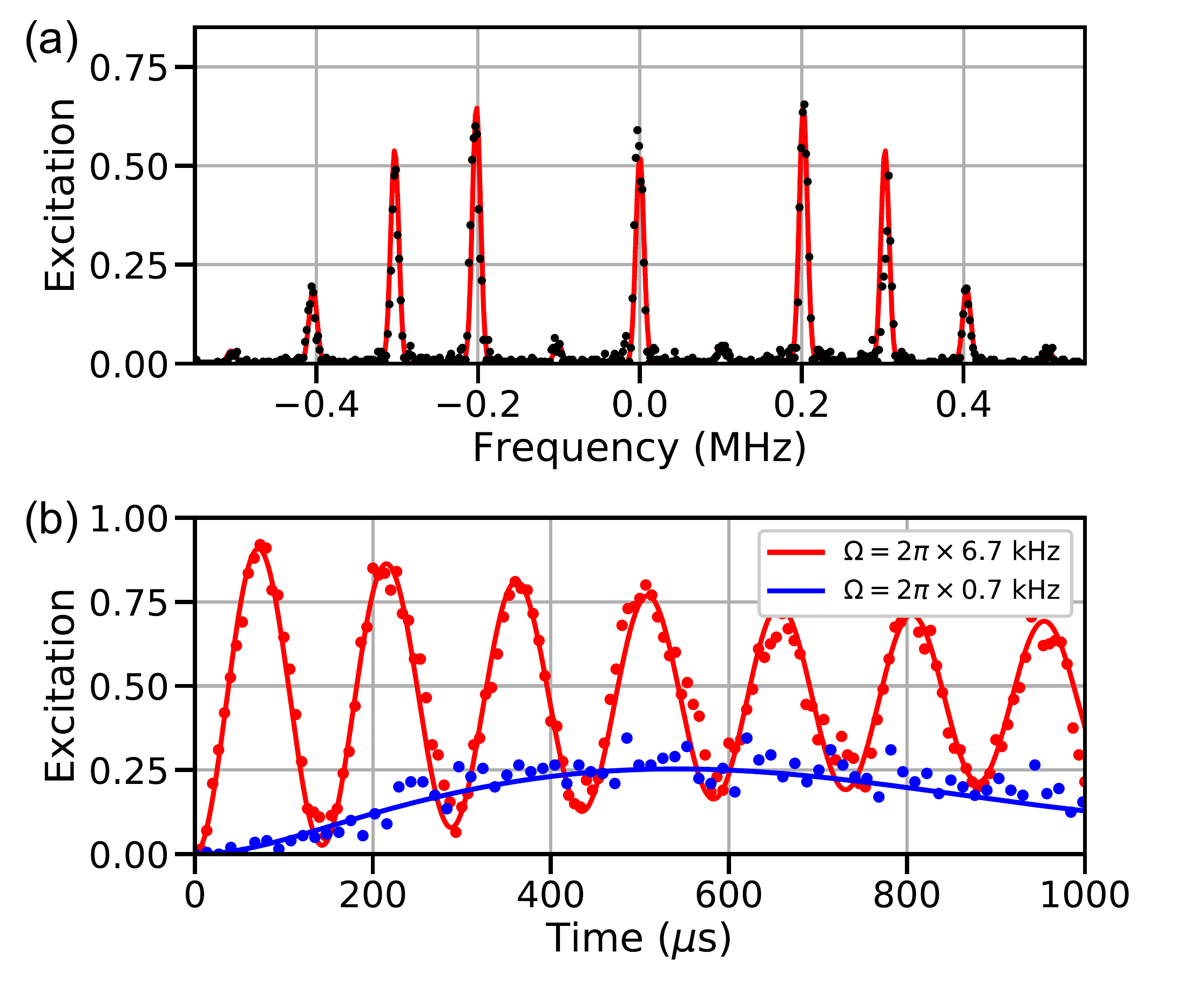}
    \caption{(a) Measured spectrum of the ion crystal prepared at 100~kHz. The theory curve is obtained by fitting for the laser angle using equation \ref{eq:mat_el} and an empirical rotation frequency of 101~kHz. (b) Rabi oscillations performed on the fourth order angular momentum sideband of a 100~kHz rotating crystal. The two curves are measured with different laser powers but the same rotation preparation. The data is fit for Rabi frequency, given by $\Omega$ in the legend, and a shared $\sigma_\ell = 45.6\pm1.0$.}
    \label{fig:couplings}
\end{figure}

A spectrum around the $|S\rangle \rightarrow |D\rangle$ transition (Fig.~\ref{fig:couplings}a) after rotation preparation shows sidebands at integer multiples of 101~kHz, a 1\% offset from our target frequency $f_\textrm{rot}$. We believe this offset is due to a small diabaticity in the spin-up process. We fit the spectrum with the coupling strengths from Eq.~\ref{eq:mat_el} and see that this spectrum is consistent with the addressing laser positioned at an angle of $\theta=82.4^\circ$ with respect to the rotor plane, where $\Delta\ell_\textrm{max} \approx 4$. 

The magnitude of the coupling matrix element between angular momentum states $\ket{\ell}$ and $\ket{\ell+\Delta\ell}$ (Eq.~\ref{eq:mat_el}), and therefore the Rabi frequency, is independent of $\ell$, while the energy splitting (and hence the spectral sideband frequency) does depend on $\ell$. Therefore, addressing rotational sidebands beginning from superpositions or mixtures of many angular momentum eigenstates yields a sum of Rabi oscillations with slightly different detunings:
\begin{equation}\label{eq:Rabi}
    P(\text{D}) = \sum_\ell \abs{c_{\ell}}^2 \frac{\Omega^2}{\Omega^2+\delta_\ell^2} \sin[2](\frac12\sqrt{\Omega^2+\delta_\ell^2}\;t)
\end{equation}
Here, $P(\text{D})$ is the probability of the ions individually being excited to the $\ket{D}$ state, $\abs{c_{\ell}}^2$ is the initial population of the state $\ket{\ell}$, $\Omega$ is the resonant Rabi frequency, $\delta_\ell = 2\omega_{\rm r}(\ell_0-\ell)\Delta\ell$ is the detuning of the transition from $\ket{\ell}$ assuming the laser is on resonance with the center of the distribution $\ell_0$, and $t$ is the coupling time. Taking $\abs{c_{\ell}}^2$ to be Gaussian distributed implies a spectral linewidth of $\gamma_\ell = 4\omega_{\rm r}\sigma_\ell\Delta\ell$. As a result, the functional form of the Rabi oscillations given by Eq.~\ref{eq:Rabi} will depend on the relative values of $\gamma_\ell$ and $\Omega$. If $\Omega \gg \gamma_\ell$, then all $\ket{\ell}\to\ket{\ell+\Delta\ell}$ transitions are nearly resonant and Rabi oscillations can be observed with high contrast. Otherwise, there are significant contributions from detuned transitions and the contrast of the Rabi oscillation is reduced. This reduction of contrast is increasingly problematic with increasing $\Delta\ell$ as the spread of the line in frequency space increases. This dependence allows us to infer $\sigma_\ell$ by fitting Rabi oscillations to Eq.~\ref{eq:Rabi}. Fig.~\ref{fig:couplings}b shows Rabi oscillations for the same state preparation (constant $\sigma_\ell$) with two different values of $\Omega$ for $\Delta\ell=4$. For the blue curve, $\Omega = 2\pi\times 0.7 \textrm{ kHz}< \gamma_\ell  = 2\pi\times 4.4$~kHz. The contrast of the oscillations is about 25\% at the $\pi$ time and saturates well below 0.5 excitation. However, with the same preparation, the red curve with $\Omega =2\pi\times6.7$~kHz shows we can achieve oscillations with over 90\% contrast on fourth order sidebands. The theory matches the data well for the fit value of $\sigma_\ell = 45.6\pm1.0$.

The high-contrast state manipulation demonstrated above allows us to create and probe the coherence of angular momentum superposition states with Ramsey interferometry. Driving a $\pi/2$ pulse on a sideband $\Delta\ell$ from an initial state $\ket{\Psi(0)} = \sum_\ell c_\ell\ket{SS,\ell}$ and waiting for a time $t$ produces the following superposition:
\begin{equation} 
    \begin{split}
        \ket{\Psi(t)} = \frac12 \sum_\ell c_\ell[&\ket{SS,\ell} \\
        &+ e^{-2i\omega_{\rm r}(\ell-\ell_0)\Delta\ell t}\ket{SD,\ell+\Delta\ell} \\ 
        &\pm e^{-2i\omega_{\rm r}(\ell-\ell_0)\Delta\ell t} \ket{DS,\ell+\Delta\ell}\\
        &+ e^{-4i\omega_{\rm r}(\ell-\ell_0)\Delta\ell t}\ket{DD,\ell+2\Delta\ell}]
    \end{split}
    \label{eq:ramsey}
\end{equation}

\noindent
Here we keep track of phases only to first order in $\Delta\ell/\ell$ in the rotating wave approximation and we have assumed perfect $\pi/2$ pulses at laser frequency resonant with the transition in the center of the distribution. This is a sum over individual manifolds, each corresponding to a single initial angular momentum eigenstate that acquires phase at its own rate. Eq.~\ref{eq:ramsey} assumes the superposition $\ket{\Psi(0)}$ is a pure state though a mixed state would demonstrate the same dynamics.

\begin{figure}
    \centering
    \includegraphics[scale=.37]{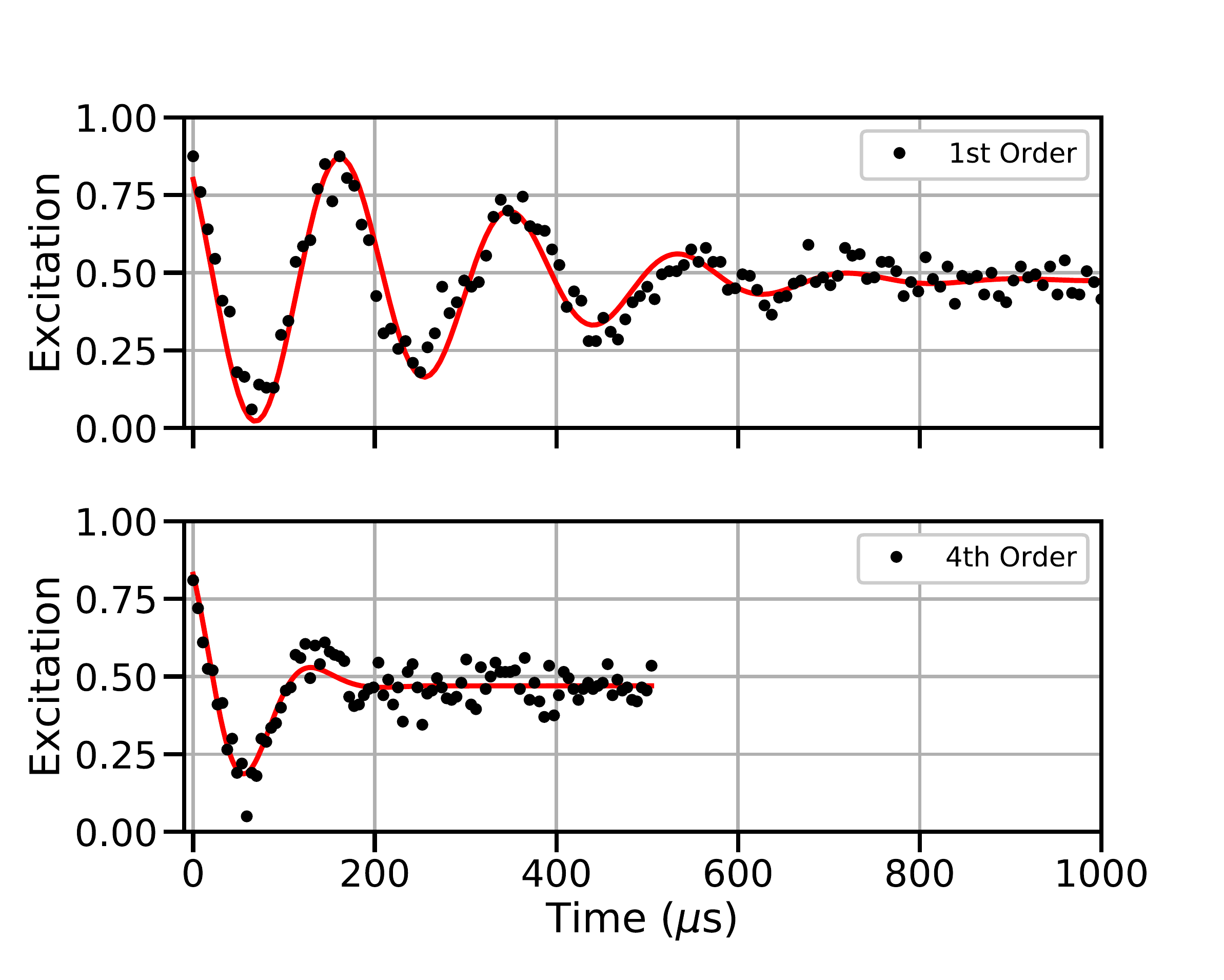}
    \caption{Ramsey experiment on angular momentum sidebands for $\Delta\ell = 1$ and $\Delta\ell = 4$ with an overall detuning of 6~kHz. Fits are made with a single $\sigma_\ell$ for both curves and individual detunings and Rabi frequencies for each curve as free parameters.} 
    \label{fig:ramsey}
\end{figure}

If we apply a second $\pi/2$ pulse after time $t$, we expect a loss of contrast at rate $\gamma_\ell$ due to the width of the line in frequency space as the phase evolution of each superposition beats against one another. Fig.~\ref{fig:ramsey} shows Ramsey experiments on the first and fourth order rotational sidebands for the same state preparation. We fit these curves to extract $\sigma_\ell$ and the overall detuning $\Delta$ with no ad-hoc decay factor included. The decay is predicted only from the beating of different manifolds of angular momentum eigenstate superpositions against each other. As expected, the fourth order superposition dephases four times as quickly as the first order. Fitting the data returns $\gamma_\ell/\Delta \ell = 1.10\pm0.03$~kHz and agrees well for both curves. This corresponds to  $\sigma_\ell = 42.7\pm1.3$ which is also similar to the state distribution observed in the Rabi oscillations. The dephasing can also be intuitively understood in the spatial domain. After driving a $\pi/2$ pulse, two branches of an interferometer exist where one branch is rotating $\Delta\ell \times 12.9$~Hz faster than the other. Therefore, once the ions have traveled far enough to become spatially separated, the contrast vanishes as they are no longer able to interfere spatially.

There are many applications for coherent control over the angular momentum mode of trapped ion systems. For example, the unique phase evolution between angular momentum states of a trapped ion crystal can be exploited for sensing and tests of fundamental physics axioms. For example, at a Ramsey time $t_\textrm{revival}=\frac{\pi}{\omega_{\rm r}\Delta\ell}$, the phase of each superposition becomes an integer multiple of $2\pi$, manifesting as a revival in the contrast of the excitation. Taking advantage of the symmetrization requirement to create a fully odd or even rotational mode occupation under particle exchange would induce an additional revival at $t_\textrm{revival}/2$, which would demonstrate the indistiguishability of the two \Ca ions even as they are separated by 6.27~$\mu$m at all times~\cite{Dieks2011, Roos2017}.

Moreover, if the rotation frequency could be made comparable to the trap frequency, the control techniques presented here could be used to study the regime in which the rotational mode of the ion crystal is strongly coupled to the stretch mode through the Coriolis force, allowing the study of the rotational dynamics of lightly bound molecules~\cite{Laurie1964,Yuan2011}. Currently, the rotation frequency is limited to a few hundred kHz by internal electronic filtering.  Additionally, the final occupation of angular momentum states is limited by the release step of the rotational mode preparation sequence. With a high precision voltage source, the ramping of the quadrupole potential could in principle be optimized to prepare the system in an angular momentum eigenstate.

In this paper, we have implemented a protocol for controlling the rotational degree of freedom in a symmetric ring ion crystal. Though this work was performed with two ions, the methods and results presented extend naturally to larger system sizes. By preparing the system in a high angular momentum state, we spectroscopically isolate transitions that selectively change the rotational quantum state. We demonstrate that angular momentum transitions are coherent and that their behavior agrees well with theory. This demonstrates the the basic control one needs to add rotational states to the toolbox available to the trapped ion community. With control over these shared motional states, we can now consider more complex Hamiltonian engineering, simulation of more diverse systems, and new tests of fundamental physics.

\begin{acknowledgments}
This work has been supported by ONR through Grant
No. N00014-17-1-2278 and by the NSF Grant No. PHY
1620838. E. U. acknowledges support by the NSF Graduate Research Fellowship under Grant No. 1106400. The authors would additionally like the recognize the help of Lorenzo Leandro.
\end{acknowledgments}

\bibliographystyle{apsrev4-1}

\bibliography{references}
\end{document}